\documentclass[aps,prl, amsmath, superscriptaddress, twocolumn,sort&compress,floatfix, amssymb]{revtex4}

\usepackage{times}
\usepackage{amsmath,amsfonts,amssymb}
\usepackage{wrapfig}
\usepackage{graphicx}
\usepackage{ulem}
\usepackage{braket}
\usepackage{color}

\begin{document}

\author{Elmar Haller}
\author{Mattias Gustavsson}
\author{Manfred J. Mark}
\author{Johann G. Danzl}
\author{Russell Hart}
\affiliation{Institut f\"ur Experimentalphysik and Zentrum f\"ur Quantenphysik, Universit\"at Innsbruck, Technikerstra{\ss}e 25, 6020 Innsbruck, Austria}
\author{Guido Pupillo}
\affiliation{ Institut f{\"u}r Theoretische Physik, Universit{\"a}t Innsbruck, Technikerstra{\ss}e 25, A--6020 Innsbruck,Austria}
\affiliation{Institut f\"ur Quantenoptik und Quanteninformation der \"{O}sterreichischen Akademie der Wissenschaften, Technikerstra{\ss}e 21a}
\author{Hanns-Christoph N\"agerl}
\affiliation{Institut f\"ur Experimentalphysik and Zentrum f\"ur Quantenphysik, Universit\"at Innsbruck, Technikerstra{\ss}e 25, 6020 Innsbruck, Austria}

\title{Realization of an Excited, Strongly-Correlated Quantum Gas Phase}

\begin{abstract}
  Ultracold atomic physics offers myriad possibilities to study strongly correlated many-body systems in lower dimensions. Typically, only ground state phases are accessible. Using a tunable quantum gas of bosonic cesium atoms, we realize and control in one dimensional geometry a highly excited quantum phase that is stabilized in the presence of attractive interactions by maintaining and strengthening quantum correlations across a confine\-ment-induced resonance. We diagnose the crossover from repulsive to attractive interactions in terms of the stiffness and the energy of the system. Our results open up the experimental study of metastable excited many-body phases with strong correlations and their dynamical properties.
\end{abstract}

\maketitle

In many-body quantum physics the interplay between strong interactions and confinement to a low-dimensional geometry amplifies the effects of quantum fluctuations and correlations. A remarkable example in one dimension is the Tonks-Girardeau (TG) gas, where bosons with strong repulsive interactions minimize their interaction energy by avoiding spatial overlap and acquire fermionic properties \cite{Girardeau1960,Lieb1963}. Evidence for this ground state phase was found using Bose-Einstein condensates (BEC) loaded into optical lattices \cite{Paredes2004,Kinoshita2004}. While many-body quantum systems are usually found in their ground state phases, long-lived excited state phases are responsible for some of the most striking physical effects, examples ranging from vortex lattices in superfluids to subtle topological excitations in spin liquids~\cite{Alet2006}. However, the experimental realization of excited phases is difficult, as these usually quickly decay by intrinsic effects or by coupling to the environment. In this context, cold atoms \cite{Petrov2000,Moritz2003,Paredes2004,Kinoshita2004,Tolra2004,Petrov2004,Hofferberth2007,Syassen2008,Bloch2008} may provide unique opportunities for the realization of long-lived, strongly interacting, excited many-body phases due to the excellent decoupling from the environment and the tunability of interactions via, for example, Feshbach resonances.

For an ultracold one-dimensional (1D) system of bosons, we prepare a highly-excited many-body phase known as the super-Tonks-Girardeau (sTG) gas \cite{Astrakharchik2005}. In this highly-correlated quantum phase, interactions are attractive, and rapid decay into a cluster-type ground state is in principle possible. However, a surprising property of this many-body phase is its metastability. Attractive interactions strengthen correlations between particle positions and ensure, similar to an effective long-range repulsive interaction, that particles rarely come together. To realize this exotic phase, we observe and exploit a 1D confinement-induced resonance (CIR) \cite{Bergeman2003,Olshanii1998}. This resonance allows us to first enter deeply into the repulsive TG regime to establish strong particle correlations and then to switch interactions from strongly repulsive to strongly attractive. The frequency ratio of the two lowest-energy collective modes \cite{Menotti2002} provides accurate diagnostics for the crossover from the TG to the sTG regime. In particle loss and expansion measurements we study the time evolution of the system through the crossover.

\begin{figure}[t]
\centering
\includegraphics[width=0.5\textwidth]{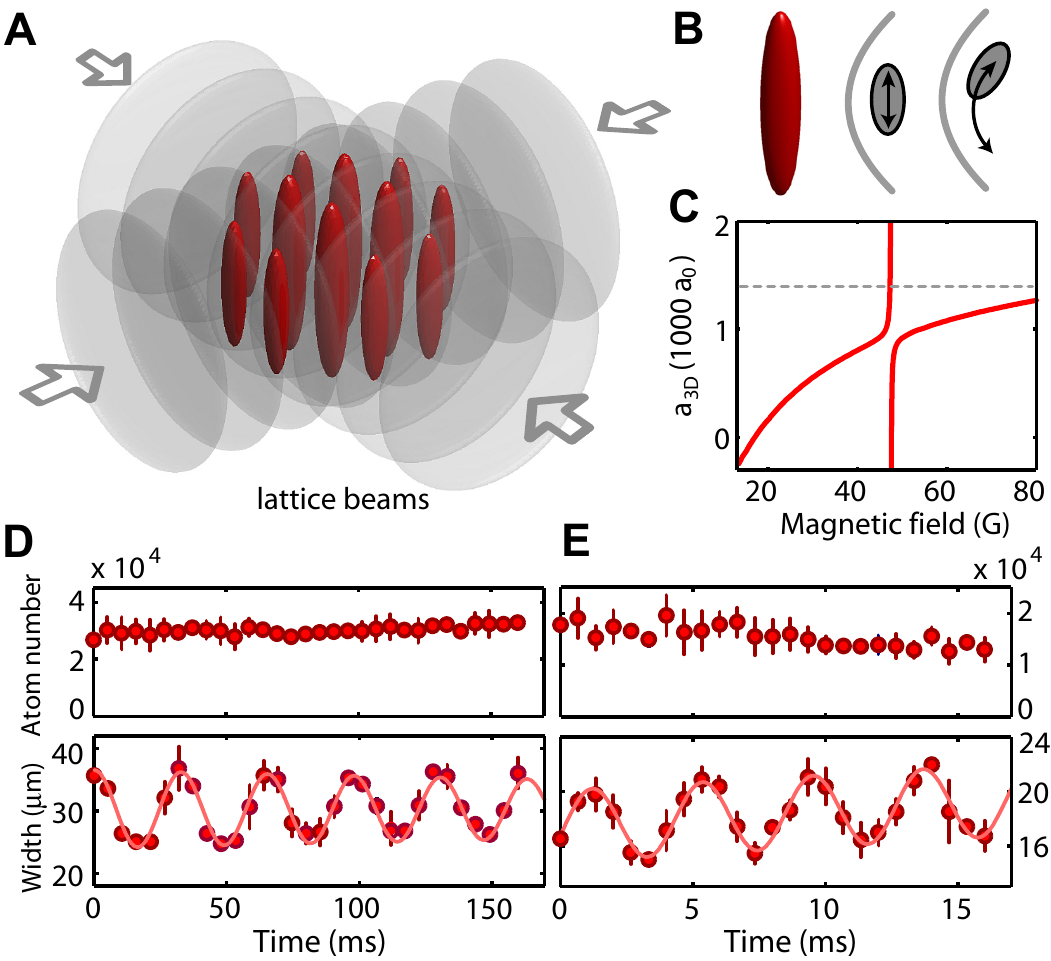}
\caption{{\textbf A}, Experimental setup. The lattice potential is created by two retro-reflected laser beams confining the atoms to an array of one-dimensional tubes with equipotential surfaces shown in red.
{\textbf B}, Along each tube (left) we excite the lowest compressional mode (center) and compare its frequency to the dipole mode (right). {\textbf C}, The strength of the interatomic interaction is adjusted by tuning the s-wave scattering length $a_\mathrm{3D}$. The background scattering length rises gently from $0$ to $1240$ a$_0$ when the magnetic field $B$ is tuned from $17$ to $76$ G. Further tuning is possible near a Feshbach resonance at $47.78(1)$ G to absolute values beyond $4000$ a$_0$. The dashed line indicates $a_{\perp}/C$ for a transversal trap frequency of $\omega_{\perp}=2\pi \times 13.1$ kHz. {\textbf D} and {\textbf E} present typical data sets for the compressional mode in the TG and sTG regime at $a_\mathrm{3D} = 875(1)$ a$_0$ and $a_\mathrm{3D}= 2300(200)$ a$_0$, respectively. The upper panels show the atom number, the lower panels show the $1/e$-cloud-width after time-of-flight. The solid lines in the lower panels are sinusoidal fits (see online material), yielding the oscillation frequencies $\omega_C = 2 \pi \times 30.6(3) $ Hz and $\omega_C = 2 \pi \times 241(1) $ Hz, respectively.}
\end{figure}

We tune the strength of the interaction as characterized by the three-dimensional (3D) scattering length $a_\mathrm{3D}$  by means of a magnetically-induced Fesh\-bach resonance \cite{Inouye1998}. For a 1D system, a CIR arises and strongly modifies the 1D scattering properties when $a_\mathrm{3D}$ approaches the harmonic oscillator length $a_\perp=\sqrt{\hbar/(m \omega_\perp)}$ of the transversal confinement with trap frequency $\omega_\perp$ \cite{Olshanii1998,Bergeman2003}. Here, $m$ is the mass of the particles and $\hbar$ is Planck's constant divided by $2\pi$. More precisely, the coupling constant $g_\mathrm{1D}$ of the 1D $\delta$-function contact potential $U_\mathrm{1D}(z) = g_\mathrm{1D} \delta(z)$ behaves as \cite{Bergeman2003}
\begin{equation}
g_\mathrm{1D} = -\frac{2 \hbar^2}{m a_\mathrm{1D}} = \frac{2 \hbar^2 a_\mathrm{3D}}{m a_\perp^2} \frac{1}{1-C \ a_\mathrm{3D}/a_{\perp}} \label{g1D},
\end{equation}
where $a_\mathrm{1D}$ is the 1D scattering length defined by this equation and $C=1.0326$ is a constant. Thus, the CIR allows tuning of $g_\mathrm{1D}$. For values of $a_\mathrm{3D}$ less but close to $a_{\perp}/C$ ($ a_\mathrm{3D}\lesssim a_{\perp}/C$) the coupling parameter $g_\mathrm{1D}$ is large and positive, and for $a_\mathrm{3D} \gtrsim a_{\perp}/C$ it is large and negative, leading to an effectively attractive interaction. For homogenous systems with $g_\mathrm{1D}>0$, it is customary to characterize the strength of interactions by the Lieb-Liniger parameter $\gamma=g_{\rm 1D} m /( \hbar^2 n_{\rm 1D})$, where $n_{\rm 1D}$ is the linear 1D density of the system \cite{Lieb1963,Petrov2000}. The TG gas corresponds to the limit $\gamma \gg 1$ or $g_{\rm 1D} \! \to \! \infty$. As interactions are increased, the system becomes strongly correlated and is fully dominated by its kinetic energy. In previous experiments, without the capability to tune $a_\mathrm{3D}$, a maximum of $\gamma \approx 5.5$ was achieved \cite{Kinoshita2004}, while an effective strength $\gamma_{\rm eff}\approx 200$ was reached with an additional shallow lattice potential along the longitudinal direction \cite{Paredes2004}. In the former experiment, a saturation for the size and energy of the 1D system was observed, and in the latter experiment the momentum distribution was studied.

\begin{figure}[t]
\centering
\includegraphics[width=0.5\textwidth]{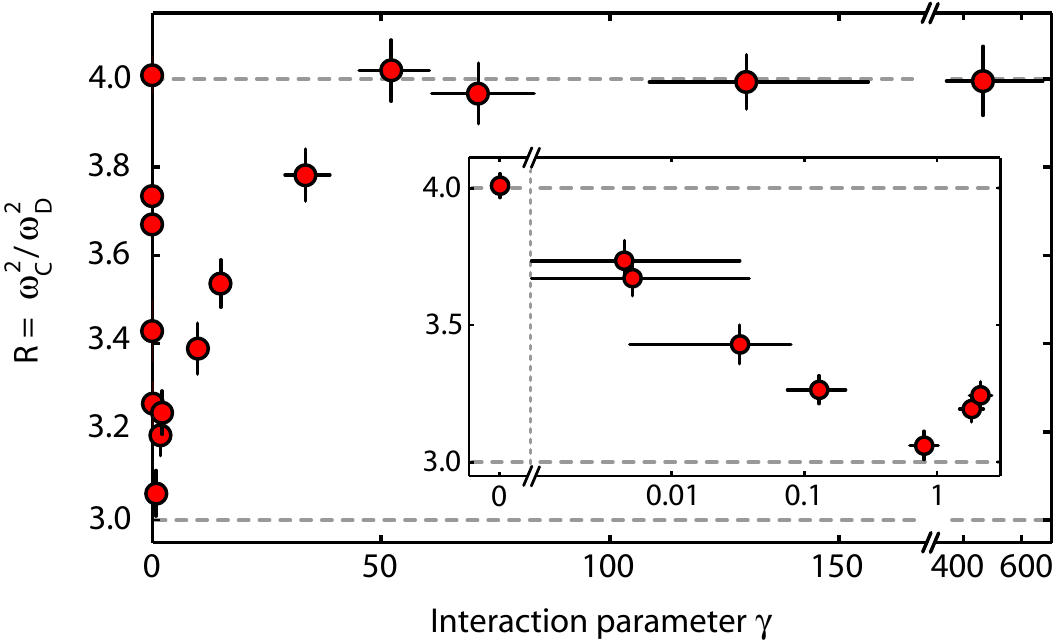}
\caption{
Transition from the non-interacting regime via the mean-field TF regime into the TG regime. The squared frequency ratio $R=\omega_C^2/\omega_D^2$ of the lowest compressional mode with frequency $\omega_C$ and the dipole mode with frequency $\omega_D$ serves as an indicator for the different regimes of interaction. For increasing interactions from $\gamma=0$ to $\gamma \approx 500$ the system passes from the ideal gas regime ($R=4$) to the 1D TF regime ($R \approx 3$) and then deeply into the TG regime ($R=4$). The inset shows the transition from the non-interacting regime to the mean-field regime in more detail. The vertical error bars refer to standard error and the horizontal error bars reflect the uncertainty in determining $a_\mathrm{1D}$ and $n_\mathrm{1D}$ (see online material). The horizontal error bar on the data point at $\gamma=0$ (not shown in the inset) is $\pm 0.03$ a$_0$.}
\end{figure}

But what happens in the case of strong attractive interactions $g_\mathrm{1D} \! \to \! -\infty$, i.e. $a_\mathrm{1D} \gtrsim 0$? The ground state for a system of $N$ attractively interacting bosons in 1D is a cluster state \cite{McGuire1965,Tempfli2008}, which one would expect, in a cold atom system, to decay quickly via molecular channels. However, by crossing the CIR from the TG side, i.e. switching interactions from $g_\mathrm{1D}=+\infty$ to $g_\mathrm{1D}=-\infty$, an excited gas-like phase, the sTG gas, should be accessible \cite{Astrakharchik2005}. Is this excited phase stable, i.e. does it exist at all? The expectation is that the large kinetic energy inherited from the TG gas, in a Fermi-pressure like manner, prevents the gas from collapsing \cite{Batchelor05}. This stability can most simply be inferred from a Bethe-ansatz solution to the Lieb-Liniger model with attractive interactions \cite{Astracharchik2004,Batchelor05}. This ansatz yields for the energy per particle $E/N \approx \hbar^2 \pi^2 n_{\rm 1D}^2/[6 m (1-n_{\rm 1D} a_{\rm 1D})^2]$, corresponding to the energy of a gas of hard rods \cite{Girardeau1960}, for which $a_{\rm 1D}$ represents the excluded volume. This results in a positive inverse compressibility and also in an increased stiffness of the systems as long as $n_{\rm 1D} a_{\rm 1D}$ is sufficiently small. Interestingly, in this phase the density correlations are even stronger than in the TG gas, as they show a power-law decay that is slower than for a TG gas \cite{Astrakharchik2005}, indicating an effective long-range interaction.

We realize the crossover all the way from a non-interacting gas via the 1D mean-field Thomas-Fermi (TF) regime to a TG gas and then to a sTG gas. We exploit the fact that our 1D systems possess weak harmonic confinement along the axial direction characterized by the confinement length $a_\parallel$. Whereas the frequency $\omega_D$ of the lowest dipole mode depends only on the confinement, the frequency $\omega_C$ of the lowest axial compressional mode is sensitive to the various regimes of interaction \cite{Menotti2002}. For the non-interacting system one expects $R\equiv\omega_C^2/\omega_D^2=4$. This value then changes to $R=3$ for weakly repulsive interactions in a 1D TF regime \cite{Moritz2003}. For increasing positive interaction strength, $R$ is expected to change smoothly to 4 when entering the TG regime as the system becomes fermionized and hence effectively non-interacting. A rise beyond the value of 4, after crossing the CIR, would then constitute clear evidence for the sTG regime \cite{Astrakharchik2005}. As $a_\mathrm{1D}$ is further increased, the system will finally become unstable and $ R $ is expected to turn over and drop towards zero. For a harmonically confined system, the point of instability is reached when the overall length of the system of hard rods, $N a_\mathrm{1D}$, becomes of the order of the size $\sqrt{N} a_{\parallel}$ for the wave function of N non-interacting fermions, i.e. $A \equiv N a_\mathrm{1D}/(\sqrt{N} a_{\parallel}) \approx 1$. We use $A^2$ as an alternative parameter to $\gamma$ to characterize the strength of the interaction as it accounts for the harmonic confinement.

\begin{figure}[t]
\centering
\includegraphics[width=0.5\textwidth]{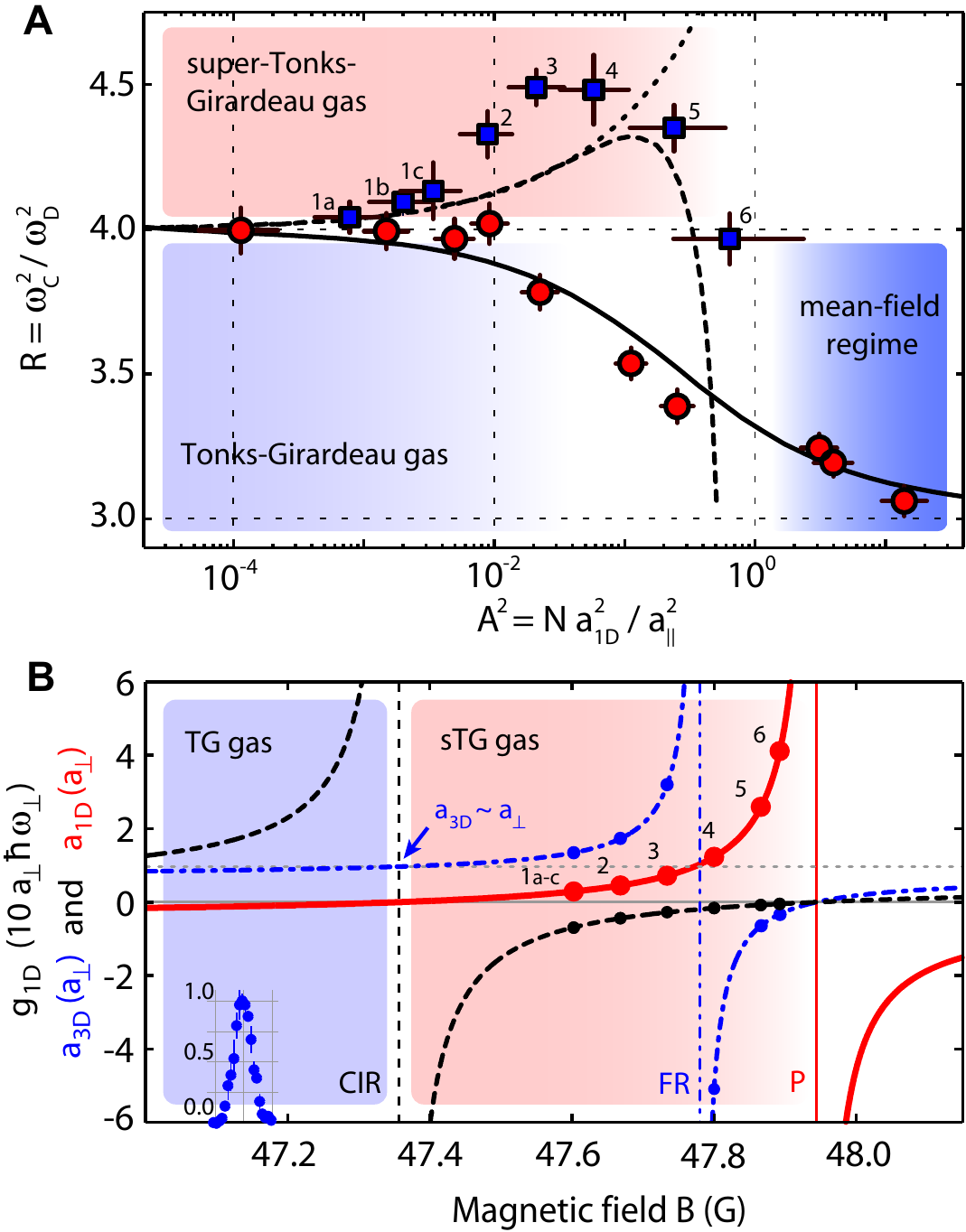}
\caption{{\textbf A}, The ratio $R=\omega_C^2/\omega_D^2$ is plotted as a function of the interaction parameter $A^2 = N a_\mathrm{1D}^2/a_{\parallel}^2$. The squares show the measurements in the attractive regime ($g_\mathrm{1D}<0$), providing evidence for the super-Tonks-Girardeau gas. The circles show the transition from the TF to the TG regime ($g_\mathrm{1D}>0$, same data as in Fig. 2 for $\gamma>1$). The solid (dashed) line presents the theoretical data for $g_\mathrm{1D}>0$ ($g_\mathrm{1D}<0$) by Astrakharchik et al.\cite{Astrakharchik2005}. The dotted line corresponds to the model of hard rods. For reference, the measurements for $g_\mathrm{1D}<0$ are numbered. Data points 1c to 6 are taken at $ \omega_D = 2\pi \times 115.6(3)$ Hz. For data points 1a and 1b the trap frequency is $ \omega_D = 2\pi \times 22.4(1)$ Hz and $ \omega_D = 2\pi \times 52.3(1)$ Hz, respectively. For all measurements in the sTG regime $a_\perp=1346(5)$ a$_0$. {\textbf B}, The parameters $a_\mathrm{3D}$ (dashed-dotted),  $a_\mathrm{1D}$ (solid), and $g_\mathrm{1D}$ (dashed) are plotted in the vicinity of the Feshbach resonance (FR) at $47.78(1)$ G. The horizontal dotted line indicates the value of $a_\perp/C$. The pole of the CIR is at $47.36(2)$ G. $a_\mathrm{1D}$ has a pole (P) at $47.96(2)$ G. The bell-shaped curve at the bottom left indicates the atomic distribution as a function of the magnetic field determined from high-resolution microwave spectroscopy.}
\end{figure}

We start from a 3D Bose-Einstein condensate (BEC) with up to $2 \times 10^5$ Cs atoms with no detectable thermal fraction in a crossed-beam dipole trap with magnetic levitation \cite{Weber2003}. Depending on the interaction regime to be studied, we then set the number of atoms in the BEC to values in the range of $(1-4) \times 10^4$ by means of forced radio-frequency evaporation. To confine the atoms in 1D, i.e. to freeze out transversal motion, we use a two-dimensional optical lattice \cite{Bloch2008}, which forms an array of vertically oriented elongated tubes with an aspect ratio that we set to values between $100$ and $1000$ (Fig. 1A). We occupy between $3000-6000$ independent tubes with $8$-$25$ atoms in the center tube. The interaction strength $g_\mathrm{1D}$ is controlled by magnetic tuning of $a_\mathrm{3D}$ by means of a combination of a broad and a narrow Feshbach resonance (Fig. 1C) with poles at $B=-11.1(6)$ G and $B=47.78(1)$ G and widths of about $29.2$ G and $164$ mG, respectively \cite{Lange2009}. The broad resonance provides a slow variation of $a_\mathrm{3D}$, allowing us to gently tune $a_\mathrm{3D}$ from $0$ a$_0$ near $17.119$ G to about $1240$ a$_0$ near $76$ G, while the narrow resonance allows us to tune $a_\mathrm{3D}$ to absolute values beyond $4000$ a$_0$ given our magnetic field control. We convert the applied magnetic field $B$ into $a_\mathrm{3D}$ using the fit formula of Ref. 23. A magnetic field gradient, used to levitate the atomic sample \cite{onlinematerial}, introduces a small spread in the value of $a_\mathrm{3D}$ across the sample.

To determine the oscillation frequencies $\omega_C$ and $\omega_D$ of the fundamental modes (Fig. 1 B), we excite each mode separately at a given value of the magnetic field $B$ \cite{onlinematerial} and let the atoms evolve for a varying amount of hold time. The distribution is then imaged in momentum
space by taking an absorption picture after release and expansion. To avoid possible broadening effects due to interaction during the initial expansion, $a_\mathrm{3D}$ is set to zero near $B=17.119$ G at the moment of release. To extract the frequency, we determine for each hold time the axial $1/e$-width of the distribution and then fit a damped sinusoid with linear offset to this data. Typical measurements of $\omega_C$ are shown in Fig. 1 D and E. Whereas the atom number remains constant for $g_\mathrm{1D}>0$, we observe some atom loss and a slight broadening of the distribution for attractive 1D interactions. In all parameter regimes, the 1D system is sufficiently stable to allow a reliable measurement of $\omega_C$.

First, we show that we can tune the system from the non-interacting regime deeply into the repulsive TG regime (Fig. 2). In agreement with expectations, the value for $R=\omega_C^2/\omega_D^2$ first drops from 4 to 3 and then increases back to 4 as $\gamma$ is tuned by means of the gently-varying background scattering length. We find that the TG regime is fully reached for $\gamma>50$. A further increase to values up to $ \gamma \approx 500$ does not lead to changes for $R$. Note that, as $a_\mathrm{3D}$ approaches $a_\perp$, the divergence of $g_{\mathrm{1D}}$ according to Eq. 1 has to be taken into account when determining $\gamma$ \cite{onlinematerial}. Heating of the system can be excluded as we can return to a 3D BEC without significant thermal background when ramping down the lattice potential.

\begin{figure}[t]
\centering
\includegraphics[width=0.5\textwidth]{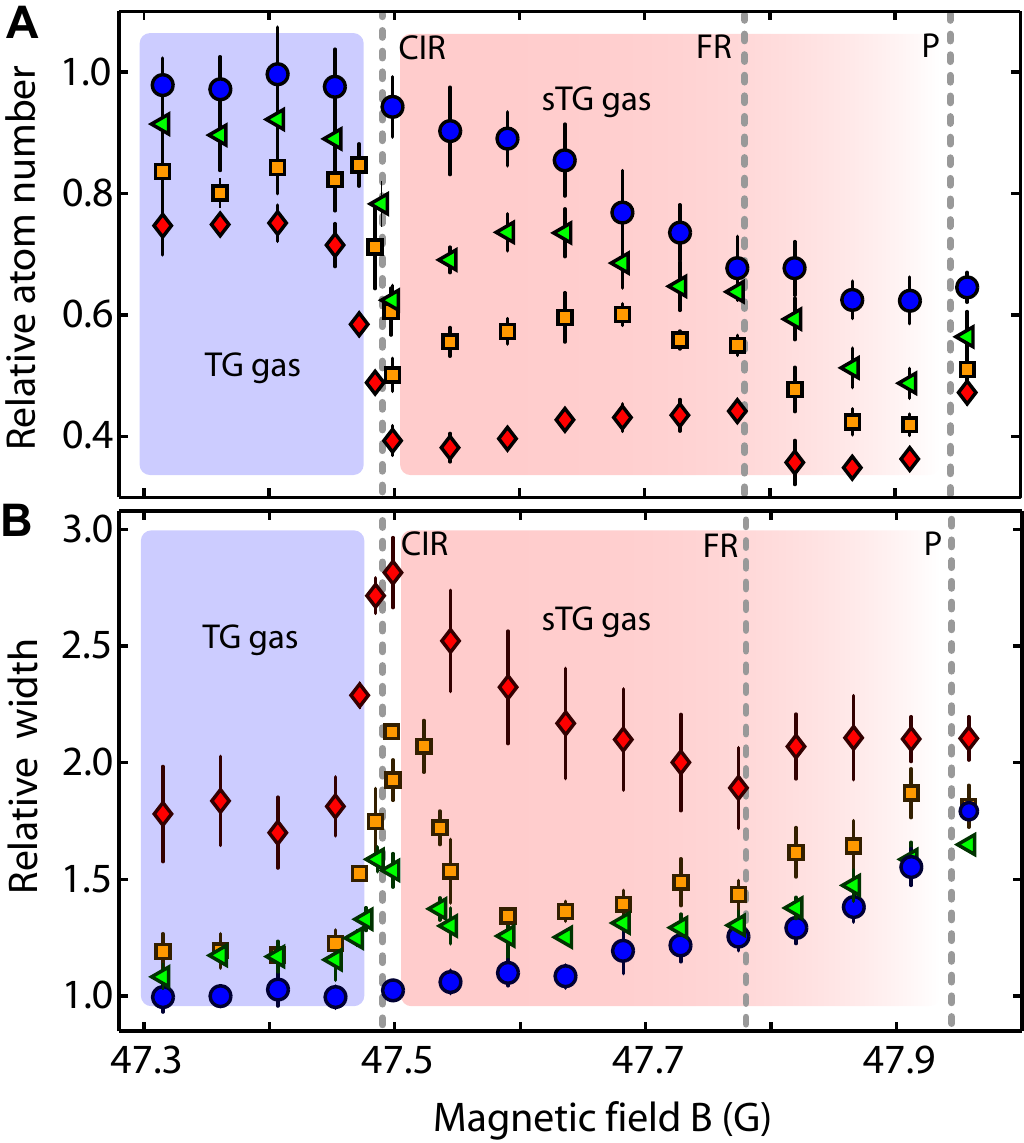}
\caption{ Stability and kinetic energy in the TG and sTG regimes.
{\textbf A}, relative number of atoms remaining and {\textbf B}, relative $1/e$-width along the axial direction after $10$ ms expansion, after a hold time $\tau=$ $10$, $50$, $100$, and $200$ ms (circles, triangles, squares, and diamonds, respectively) at a given magnetic field $B$. The position of the CIR, the pole of the Feshbach resonance (FR), and the pole for $a_\mathrm{1D}$ (P) are as indicated. For these measurements $a_\perp=1523(6)$ a$_0$ and $ \omega_D = 2\pi \times 115.6(3)$ Hz. The atom number is normalized to the initial value of $1.7(1) \times 10^4$ and the width is normalized to the initial value in the TG regime.}
\end{figure}

The attractive regime is entered by crossing the CIR on the low-field wing of the 47.78 G Feshbach resonance. $ a_\mathrm{1D} $ is now small and positive. The central results of this work are summarized in Fig. 3A and compared to the theoretical work of Ref. 13. We plot $R=\omega_C^2/\omega_D^2$ as a function of the interaction parameter $A^2$. For reference, Fig. 3B plots $a_\mathrm{3D}$,  $a_\mathrm{1D}$, and $g_\mathrm{1D}$ in the vicinity of the Feshbach resonance as a function of the magnetic field $B$. As the CIR is crossed and $ A^2 $ is increased, $R$ rises beyond the value of $4$. This provides clear evidence for the sTG regime as $R=4$ is the maximal value for bosons with repulsive contact interaction. This increase is expected from the model of a gas of hard rods, and our data initially follows the prediction from this model. However, as $ A^2 $ is increased, $R$ reaches a maximum and then starts to drop. The maximum of about $4.5$ is reached for $A^2 \approx 3 \times 10^{-2}$. The existence of the maximum is in qualitative agreement with the results obtained from Monte-Carlo simulations \cite{Astrakharchik2005}. The theoretical prediction, however, underestimates the measured $R$. This is probably due to the local density approximation, which may not be applicable to our system with low particle numbers. For comparison, the results from Fig. 2 for $\gamma \ge 1$ are shown. Note that $ \gamma \approx 500 $ corresponds to small values of $A^2  \approx 10^{-4} $. For this data, at higher particle numbers, there is  excellent agreement with the theoretical prediction (solid line) in the entire crossover from the mean-field regime to the TG regime\cite{Menotti2002}.

We study the stability of the system in the crossover from the TG to the sTG regime and find further evidence for the existence of the CIR by recording particle loss and measuring the axial width of the atomic cloud after release from the tubes. The axial width is a measure for the kinetic energy of the system as interactions are instantly switched off upon release. Similar conditions are used as for the measurements on the sTG regime presented in Fig. 3. The TG regime is entered adiabatically to avoid the excitation of collective modes. The system is prepared at $a_\mathrm{3D} = 887(1) $ a$_0$ at a magnetic field of $B=42.77(2)$ G with about $11$ atoms in the central tube. The magnetic field is then ramped to a specific value within $0.2$ ms and the sample is held at this value for a variable hold time $\tau$ from 10 to 200 ms. $a_\perp$ is set to $1523(6)$ a$_0$. The results (Fig. 4) for different hold times $\tau$ in the tubes show that, for $\tau = 10$ ms, corresponding to the timescale of the measurements in the sTG regime shown in Fig. 3, the transition from the TG to the sTG regime appears very smooth. There is essentially no particle loss when the system is deep in the TG regime and close to the CIR. The loss gradually increases in the attractive regime as one moves to larger values of $B$ and towards the pole for a$_\mathrm{1D}$. Correspondingly, the width of the sample exhibits a smooth behavior across the CIR, showing a slight increase for larger $B$. This behavior is consistent with the expectation of an increased energy in the sTG regime \cite{Astrakharchik2005}.

For longer hold times, the data for the atom number and the sample width develop distinct features at the calculated position of the CIR. Evidently, the system is in a transient state. For $\tau = 50$ ms, the number of remaining atoms shows a dip that correlates with a peak in the kinetic energy of the sample. Both features become more prominent and asymmetric for longer hold times ($\tau = 100$ and $200$ ms). Note that, in comparison, no pronounced effects are visible at the pole of the Feshbach resonance for $a_\mathrm{3D}$.
Our results must be connected to the fact that the energy spectrum of the system changes dramatically across the CIR, from the TG to the sTG regime \cite{Tempfli2008}. The system acquires a deeply lying ground state together with a family of lower lying many-body excited states, potentially opening up new decay channels. Also, the CIR strongly modifies the two-body scattering problem, making formation of confinement-induced molecules in transversally excited trap states \cite{Bergeman2003} possible.

The non-trivial time evolution observed in this system raises intriguing questions on possible coupling and decay mechanisms for strongly interacting excited many-body systems, in particular in the context of integrability of 1D systems \cite{Kinoshita2006}. Our results offer an example of the counter-intuitive effects that occur in many-body systems, and open up the possibility to study the dynamical properties of strongly-correlated systems with effective long-range interactions \cite{Bockrath1999,Steinberg08} under conditions where all parameters are tunable and, in fact, can be changed dynamically. Similar to magnetic Feshbach resonances in atomic scattering, we expect the confinement-induced resonance demonstrated here to serve as a general tool to tailor interactions in 1D and possibly also in 2D systems \cite{Petrov2000b}, allowing for the further investigation of strongly correlated phases in the context of cold atomic gases.

\vspace{3ex}

We thank S. Giorgini and C. Menotti for helpful discussions and for providing the theory curves shown in Fig. 3A. We are indebted to R. Grimm for generous support and to H. H\"affner and his group for the loan of a CCD camera. We gratefully acknowledge funding by the Austrian Ministry of Science and Research (Bundesministerium f\"ur Wissenschaft und Forschung) and the Austrian Science Fund (Fonds zur F\"orderung der wissenschaftlichen Forschung) in form of a START prize grant and by the European Union through the STREP FP7-ICT-2007-C project NAME-QUAM (Nanodesigning of  Atomic and MolEcular QUAntum Matter) and within the framework of the EuroQUASAR collective research project QuDeGPM. R.H. is supported by a Marie Curie International Incoming Fellowship within the 7th European Community Framework Programme.


\onecolumngrid
\newpage
\twocolumngrid

\section*{Materials and methods}

{\bf Lattice loading.} We produce a BEC of Cs atoms in the lowest hyperfine sublevel with hyperfine quantum numbers $F=3$ and $m_F=3$ in a crossed beam dipole trap with trap frequencies $\omega_{x,y,z} = 2 \pi \times (15, 20, 13)$ Hz, where $z$ denotes the vertical direction. The BEC is adiabatically transferred from the dipole trap to the array of tubes by exponentially ramping up the power in the lattice laser beams with waists $\sim 350 \ \mu$m within $500$ ms. The repulsive interaction causes the atoms to move radially outwards during the initial phase of the lattice loading in response to the strong local compression. We use this effect to vary the total number of tubes loaded and hence the atom number per tube by setting $a_\mathrm{3D}$ for the loading process to values between $40$ $a_0$ and $350$ $a_0$. For the data set in the repulsive regime (Fig.3A, circles), we exponentially ramp down the crossed beam dipole trap during the loading process and reach longitudinal and transversal trap frequencies of $\omega_D=2\pi \times 15.4(1)$ Hz and $\omega_{\perp}=2\pi \times 13.1(1)$ kHz with a transversal confinement length $a_\perp = 1440(6) $ a$_0$. Here, depending on the regime of interaction to be studied, the number of atoms in the central tube is set to values between 8 and 25. For the data set in the sTG regime (Fig.3A, squares) we increase $\omega_D$ to $2\pi \times 115.6(3)$ Hz to reduce the vertical extent of the sample and hence the variation of the magnetic field across the atom cloud, see below. For this, we keep the depth of the crossed beam dipole trap constant during the loading process and then ramp up the power in one of the beams within $100$ ms. In this regime we choose $\omega_{\perp}=2\pi \times 15.0(1)$ kHz, corresponding to $a_\perp =1346(5)$ a$_0$. The number of atoms in the central tube is set to values between 8 and 11.
\vspace{2ex}

{\bf Array of 1D tubes.} The atom number per tube becomes fixed once tunneling is suppressed during the loading process and can be determined by integrating a Thomas-Fermi profile along the tubes (1). The number of atoms in tube $(i,j)$ is given by
\begin{eqnarray*}
N_{i,j} &=& N_{0,0}\,\left[1-{\left(i \frac{d_\mathrm{lat}}{R_x}\right)}^2-{\left(j \frac{d_\mathrm{lat}}{R_y}\right)}^2\right]^{3/2} \\
 N_{0,0} &= &\frac{5 N_\mathrm{tot} d_\mathrm{lat}^2}{2\pi R_x R_y},
\end{eqnarray*}
where $N_\mathrm{tot}$ is the total atom number, $N_{0,0}$ is the occupation of the central tube, $d_\mathrm{lat}=\lambda/2$ is the lattice spacing at a wavelength $\lambda=1064.5$ nm, and $R_{x,y}$ are the Thomas-Fermi radii in the horizontal directions. To calculate the effective atom number per tube $N$, we average over the tubes, weighting each tube by its atom number. This procedure accounts for the fact that we measure an averaged frequency $\omega_C$, as $\omega_C$ is expected to slightly vary from tube to tube. The result for $\omega_C$ should be dominated by the more heavily occupied tubes close to the center of the array.
\vspace{2ex}

{\bf Magnetic levitation.} To hold the $(F\!=\!3, \ m_F\!=\!3)$ atoms in the vertically oriented tubes, magnetic levitation by means of a magnetic field gradient of 31.1 G/cm is applied. The gradient introduces a small field spread over the atomic sample. This sets our precision to tune the interaction strength. For the measurements in the sTG regime the distribution has a full width at half maximum (FWHM) of $30$ mG. We measure the atom distribution in the magnetic field by driving a magnetic-field-dependent microwave transition. A typical distribution is shown in the bottom left corner of Fig.3B.
\vspace{2ex}

{\bf Excitation of collective modes.} We use two different methods to excite the lowest compressional mode. For a measurement in the mean-field regime, we use a rapid change of the interaction strength to excite the oscillation. For this, we ramp the scattering length adiabatically in $100$ ms to a value that is approximately $50$ a$_0$ from the desired final value and then perform the last part of the ramp non-adiabatically. For a measurement in the sTG regime, we use an analogous method. We simply ramp sufficiently quickly, within about $5$ ms, all the way from the mean-field across the TG into the sTG regime. For the TG regime, we chose to excite the mode by compressing the cloud adiabatically with an additional dipole trap laser beam, starting the motion by rapidly ramping down the power of this beam. In all cases, we adjust the ramp speeds so that the measured oscillation amplitude is within $10$-$20\%$ of the initial cloud size. To excite the dipole mode at frequency $\omega_D$, we adiabatically lower the levitating magnetic field gradient and hence displace the cloud along the vertical direction. Quickly readjusting the gradient back to full levitation leads to excitation of the dipole oscillation.
\vspace{2ex}

{\bf Determination of $\gamma$.} We make a conservative estimate to determine the Lieb-Liniger interaction parameter $\gamma$
\[ \gamma = \frac{m\, g_{\mathrm{1D}}}{\hbar^2\, n_{\mathrm{{1D}}}} = \frac{2}{n_{\mathrm{1D}}\, |a_{\mathrm{{1D}}}|}.\] To take into account that the atom number varies according to $N_{i,j}$, we first calculate $\gamma_{i,j}$ for every tube separately. We calculate the center density for each tube both in the mean-field and in the TG regime and use the larger value to determine $\gamma_{i,j}$. We then take $\gamma$ as the weighted average over $\gamma_{i,j}$. The error in determining $\gamma$ largely comes from the determination of $a_{\mathrm{{1D}}}$, reflecting the magnetic field distribution across the sample.
\vspace{2ex}

{\bf References and Notes}
\begin{itemize}
\item[1.]
B. L. Tolra {\it et al.}, {\it Phys. Rev. Lett.} {\bf 92}, 190401 (2004).
\end{itemize}

\end{document}